\documentclass[10pt]{iopart}

\usepackage{tabularx}
\usepackage{subcaption}
\usepackage{graphicx}
\usepackage[xcdraw, table]{xcolor}

\begin{document}

\title{Estimating blood pressure trends and the nocturnal dip from photoplethysmography}

\author{Mustafa Radha\textsuperscript{1,2},
Koen de Groot\textsuperscript{1},
Nikita Rajani\textsuperscript{3},
Cybele CP Wong\textsuperscript{3},
Nadja Kobold\textsuperscript{3},
Valentina Vos\textsuperscript{3},
Pedro Fonseca\textsuperscript{1,2},
Nikolaos Mastellos\textsuperscript{3},
Petra A Wark\textsuperscript{3,4},
Nathalie Velthoven\textsuperscript{2},
Reinder Haakma\textsuperscript{2},
Ronald M Aarts\textsuperscript{1,2}}

\address{\textbf{1} Personal Health, Philips Research, Royal Philips, Eindhoven, The Netherlands\\
\textbf{2} Signal Processing Systems, Electrical Engineering, Eindhoven University of Technology, Eindhoven, The Netherlands\\
\textbf{3} Global eHealth Unit, Department of Primary Care and Public Health, Imperial College London, London, United Kingdom\\
\textbf{4} Faculty of Health and Life Sciences, Coventry University, Coventry, United Kingdom\\}

\ead{mustafa.radha@philips.com}
\vspace{10pt}
\begin{indented}
\item[]October 2018
\end{indented}

\begin{abstract}\\

\textbf{Objective:} Evaluate a method for the estimation of the nocturnal systolic blood pressure (SBP) dip from 24-hour blood pressure trends using a wrist-worn photoplethysmography (PPG) sensor and a deep neural network in free-living individuals, comparing the deep neural network to traditional machine learning and non-machine learning baselines.\\

\textbf{Approach:} A wrist-worn  PPG sensor was worn by 106 healthy individuals for 226 days during which 5111 reference values for blood pressure (BP) were obtained with a 24-hour ambulatory BP monitor and matched with the PPG sensor data. Features based on heart rate variability and pulse morphology were extracted from the PPG waveforms. Long- and short term memory (LSTM) networks, dense networks, random forests and linear regression models were trained and evaluated in their capability of tracking trends in BP, as well as the estimation of the SBP dip.\\

\textbf{Main results:} Best performance for estimating the SBP dip were obtained with a deep LSTM neural network with a root mean squared error (RMSE) of 3.12$\pm$2.20 $\Delta$mmHg and a correlation of 0.69 $(p=3*10^{-5})$. This dip was derived from trend estimates of BP which had an RMSE of 8.22$\pm$1.49 mmHg for systolic and 6.55$\pm$1.39 mmHg for diastolic BP (DBP). While other models had similar performance for the tracking of relative BP, they did not perform as well as the LSTM for the SBP dip.\\

\textbf{Significance:} The work provides first evidence for the unobtrusive estimation of the nocturnal SBP dip, a highly prognostic clinical parameter. It is also the first to evaluate unobtrusive BP measurement in a large data set of unconstrained 24-hour measurements in free-living individuals and provides evidence for the utility of LSTM models in this domain.

\end{abstract}

\section{Introduction}
\label{sec:introduction}

Cardiovascular disease is a leading cause of death worldwide \cite{roth2018global}. An important early indicator of cardiovascular risk is BP elevation known as hypertension. BP measurement is currently mainly done through point measurement (i.e. non-continuous controlled measurement at a singular point in time) using obtrusive inflatable cuffs (auscullatory and oscillatory sphygmomanometer cuffs), through which elevations in BP can be detected. Over the last decade, there has been strong evidence to suggest that BP should not be regarded as a static number, given its daily variation \cite{Hansen2010, Floras2013, Mancia2012}. In particular, the nocturnal variation in SBP has unique prognostic value. A nocturnal decrease (i.e. dip) in SBP of 5 mmHg was associated with a 17\% reduction of cardiovascular risk \cite{hermida2008chronotherapy, hermida2011decreasing, hermida2011influence}. \\

Currently, the European Society of Hypertension \cite{Mancia2013} recommends measuring the SBP dip using the ambulatory BP monitor, the current gold standard for BP measurement in daily life \cite{Pickering2005}. An ambulatory BP monitor is an oscillometric cuff worn  across a 24-hour period and that inflates at regular intervals. However the monitoring procedure is highly obtrusive, especially during sleep, as the repeated cuff inflation generates a substantial amount of noise and cuts off blood flow to the brachial artery, both of which can disturb sleep \cite{Heude45}. This makes it difficult to adopt SBP dip measurement into clinical practice.\\

This motivates the need for BP monitors that can unobtrusively and continuously measure BP, or at least the relative changes in BP, throughout the day. These relative changes, that in this manuscript are called \textit{trends}, indicate how the BP is changing from day to night without requiring to know the absolute value of BP. These diurnal variations are sufficient to extract meaningful prognostic information such as the nocturnal SBP dip. The aim of this study was to propose such a method using a wrist-worn wearable PPG sensor and evaluate it in \textit{free-living} individuals against the gold-standard ambulatory BP method. The method consists of a machine learning algorithm trained to predict how the BP changes with respect to the mean BP of a person, and then uses that information to compute the SBP dip.\\

 In Section \ref{sec:background} an overview of the existing literature on unobtrusive BP measurement is presented, current limitations are discussed and the objective and motivation of this study are given in detail. In Section \ref{sec:methods} an overview of the data collection process, the feature extraction methods, the machine learning methods and the evaluation strategies are presented. Results are presented in Section \ref{sec:results} and discussed in Section \ref{sec:discussion}.

\section{Background}
\label{sec:background}
In this section the scientific background is described. The most notable and relevant works are summarized in Table \ref{tab:state_of_art}.
\subsection{Physiological models}
Unobtrusive BP measurement research has investigated a variety of sensors and algorithms that operate non-invasively and don't require an inflatable cuff, typically through sensors that are placed on the surface of the skin. The majority of research in this area is based on the Moens-Korteweg equations, which relate the velocity of the arterial pulse wave to BP \cite{Mukkamala2015, Peter2014, Buxi2015}, which is known as the pulse arrival time (PAT). PAT is measured as the time delay between the ejection of the blood pulse from the heart (i.e. pulse onset) and the arrival of the pulse at an arterial site (i.e. pulse arrival). Most commonly, the R-peak on an electrocardiogram (ECG) is used to indicate the pulse onset while arrival is measured with PPG at peripheral sites, most commonly the wrist or finger \cite{Mukkamala2015,Muehlsteff2006,gesche2012continuous}, but also other body locations such as ears \cite{Radha2017}, ankles \cite{Radha2017}, toes \cite{Nitzan2002} the chest \cite{sola2013noninvasive}, or even using remote PPG \cite{zhang2017hybrid}. However studies show that measuring PAT in free-living individuals during the day has limited accuracy \cite{Zheng2014}. Posture compensation methods have been proposed \cite{Thomas2015,Poon2006} to improve the accuracy of day-time measurements but the effectiveness of such methods in free-living individuals remains to be assessed. Another issue with measuring PAT is obtrusiveness as it requires the placement of two sensors on separate locations on the body which is cumbersome for the individual. Thus PAT could be feasible to use in a clinical setting but not at home. Some of the above-mentioned studies were included in Table \ref{tab:state_of_art}. More comprehensive reviews of PAT-based BP were given by Mukkamala et al. \cite{Mukkamala2015} and Peter et al. \cite{Peter2014}. \\ 

Research focus then shifted towards pulse morphology analysis \cite{Elgendi2012} as a step away from PAT, inspired by physiological explanations of the dynamics in the pulse waveform in PPG. While pulse wave analysis expands the set of methods that may be used for BP estimation, these features have only been tested in controlled lab circumstances, where the effects of movement and free-living activities are not present.\\

\subsection{Machine learning models}
\label{sec:background:ml}
To avoid heavy reliance on a single physiological model for BP prediction that could be affected by confounding factors or noise, a shift is being made towards machine learning methods. This approach was pioneered by Monte-Moreno \cite{Monte-Moreno2011} who combined a set of features describing several PPG characteristics in a random forest model \cite{liaw2002classification} to predict continuous SBP. Later, similar methods were applied to continuous intensive care unit measurements \cite{Ruiz-Rodriguez2013}, providing the first evidence for continuous BP measurement using deep belief networks \cite{hinton2006fast}. Since then, similar approaches have been published that have experimented with machine learning. Sun et al. \cite{Sun2016} evaluated a linear regression model in bicycle tests for SBP prediction, combining a large number of morphological properties of the PPG. Xing et al. \cite{Xing2016} had a similar approach but used support vector regression and only evaluated on point measurement data. Miao et al. \cite{miao2017novel} used a multilayer perceptron model trained on a large intensive care database and evaluated on point measurement samples.\\ 

A recently published study compared the traditional ECG-to-PPG PAT method to machine learning models that include both PAT as well as PPG morphology in continuous measurements of 8 minutes, conducted in a lab (12 participants, 4 times each). They showed that PAT is a less accurate model in comparison to the more sophisticated linear and non-linear models such as linear regression and random forest. However the best-performing model was a deep neural network approach using LSTM cells, a type of sequence-to-sequence model \cite{sutskever2014sequence}. While normal machine learning models take as input the feature values from the sensors at one point in time and predict the BP for that same point in time \cite{Monte-Moreno2011,Ruiz-Rodriguez2013}, sequence-to-sequence models regard the classification task as mapping a sequence of inputs (sampled over time) to a sequence of outputs (sampled at the same time points). These models condition the prediction at some time $t$ not only on the feature values measured at time $t$ but also on feature values measured at other time points in the sequence. The LSTM cells achieve this through recurrence \cite{hochreiter1997long}. The prediction $h_{t}$ (at time point $t$) is based on its input $x_t$, its last prediction $h_{t-1}$ (short-term recurrence) and its internal cell state $C_t$ (long-term recurrence). The internal memory state variables are controlled through gating mechanisms. A forget gate can clear the contents of the state while an input gate can feed data into the cell state. The variables controlling the input gate, forget gate and the predictions are trainable parameters. LSTM cells are usually incorporated in neural networks and trained through back-propagation. Exact mathematical definitions are given by Hochreiter et al. \cite{hochreiter1997long} for the LSTM model and sequence-to-sequence models are extensively described by Sutskever et al. \cite{sutskever2014sequence}

\subsection{Evaluation in free-living context}
Unobtrusive approaches for the estimation of BP have most of their value in a free-living context, where users can carry on with their daily activities while their BP is measured. Such approaches could also address the estimation of the SBP dip that was introduced in the previous section.  However, although the aforementioned studies evaluated their methods using continuous data, this data was collected in the controlled environment of a lab and often across a limited time span ranging from a few minutes to half an hour. Therefore, while lab studies provide initial evidence for the feasibility of such methods, the performance reported may not extrapolate to free-living context. The only study where BP surrogates were evaluated in a free-living setting was done by Zheng et al. \cite{Zheng2014} where PAT-based BP estimation was evaluated in 12 healthy volunteers. The method's performance however, was only reported  during night-time in which person movement is minimal and yet, the reported RMSE was higher than in lab studies using PAT \cite{Muehlsteff2006,sola2013noninvasive} (see Table \ref{tab:state_of_art}).

\begin{table}
\caption{Overview of studies of unobtrusive SBP and DBP estimation, categorized by study protocol, and their obtained performance.} 
\label{tab:state_of_art}

\begin{tabular}{>{}l >{}l >{}l >{}l >{}l>{}l>{}l>{}l}
\hline
   & &  &  & \multicolumn{2}{c}{\textbf{RMSE (mmHg)}} \\ \hline \hline
  \textbf{Citation} & \textbf{Sensors} & \textbf{Method} & \textbf{N} &\textbf{SBP} & \textbf{DBP} \\ \hline \hline
  
      \multicolumn{5}{l}{\textit{Point measurements}} \\ \hline
  2011 \cite{Monte-Moreno2011} & PPG & Random forest & 410 &  $R^2: 0.91^a$ & $R^2:0.89^a$\\ 
      2016 \cite{Xing2016} & PPG & Multilayer perceptron&92  & $1.67$ & $1.29$\\ 
    2017 \cite{miao2017novel} & ECG, PPG & Support vector regression& 73 &   $3.10$ & $2.20$\\ \hline
    \multicolumn{5}{l}{\textit{Continuous measurements in controlled lab setting}} \\ \hline
  2006 \cite{Muehlsteff2006} & ECG, PPG & PAT & 18 &  $ 6.9$ & $-$\\ 
      2013 \cite{Ruiz-Rodriguez2013} & PPG & Deep belief networks & 572 &  $9.01$ & $0.47$\\ 
    2016 \cite{Sun2016} & PPG & Linear regression & 19  & 8.99 & $-^c$ \\ 
    2018 \cite{su2017predicting} & ECG, PPG & LSTM& 12 & $4.13$  & $2.8$ \\ \hline
    \multicolumn{5}{l}{\textit{Continuous measurements in 24-hr free-living setting}} \\ \hline
  2014 \cite{Zheng2014} & ECG, PPG & PAT & 10 &  $ 8.7^b$ & $-^c$
    \\ \hline
\end{tabular}	
\footnotesize{
$^a$ Performance was only reported in $R^2$\\
$^b$ Performance was only reported for night-time measurement\\
$^c$ No performance reported for DBP}\\

\end{table}

\subsection{Objective of the current study}
The aim of this study is to propose and evaluate a method using PPG morphology to estimate diurnal BP trends in free-living individuals, with the final goal of estimating the SBP dip from those trends. The model predicts \textit{relative} BP, which means that it is trained to predict change in BP with respect to the mean BP of a person. This information is sufficient to extract the SBP dip as this is also a relative measure. With a careful calibration these measurements could be used to derive absolute BP, but the calibration is not in the scope of this study. The evaluation is done using a unique data set containing 226 days of ambulatory BP measurements from 103 individuals.\\ 

The proposed method is based on a deep LSTM neural network. The performance of this proposed method is compared to more traditional methods: a random forest regressor which has no ability to infer over temporal structure, a linear regression model which in addition also has no ability to infer non-linear relations, as well as two simple baseline models for the SBP dip estimation: one using demographics and the other using heart rate.\\

This work is the first to evaluate an un-obtrusive PPG-based BP trend estimation method for the prediction of the SBP dip and is also the first to evaluate PPG morphology as a predictor of BP in free-living context. Finally, it is the first to evaluate  deep learning algorithms for BP measurement in free-living context.

\section{Methods}
\label{sec:methods}

\subsection{Data}
\label{sec:materials}
The study was conducted across three Imperial College London Healthcare NHS Trust Hospitals (Charing Cross Hospital, St. Mary's Hospital and Hammersmith Hospital) and at the Philips Research laboratory in Eindhoven, the Netherlands. The study was approved by the ethics institutions of each of the participating organisations. Participants (N=110) were healthy volunteers aged 18-65 years, who had worked in a medical setting for at least 6 months in the same type of rota, had not travelled across more than two time zones in the 30 days prior to enrolment, and who continued to work at least 21 hours a week during the course of the study. To confirm eligibility for the study, participants attended a screening visit where they had their BP taken on both arms using a standard BP monitor. Those with a difference in BP between arms of more than 10 mmHg were excluded. In addition, weight and height measurements were taken using a standard weight scale and measuring tape to determine body-mass index. Participants were excluded if the body-mass index was above 30 kg/m$^3$. Other exclusion criteria included: being pregnant, having taken BP altering medication in the last six months or the regular use of light therapy. These criteria were assessed by an intake questionnaire.\\
\\

Participants were asked to participate in up to 3 measurement days, which were each at least a week apart. At the start of each measurement day, participants were fitted with a clinically validated ambulatory BP monitor (Mobil-o-Graph NG \cite{franssen2010evaluation}) which was set up to measure BP automatically every 30 minutes. At the same time, participants were asked to wear a wrist band containing a green PPG sensor and a triaxial accelerometer. Both sensors were calibrated to sample data at 128 Hz. A more detailed description can be found in \cite{bonomi2016atrial} where the same sensors were used for the detection of atrial fibrillation. Participants were instructed on how to wear the band and at what tightness (based on arm circumference). It was decided to standardise the location of the PPG sensor in this study to the right wrist as differences in PPG are minimal across wrists \cite{buchs2005right,bozkurt2015}. Participants were asked to wear the devices for 24-hours and carry on with their regular working patterns and routine.

\begin{figure}
\includegraphics[scale=0.7]{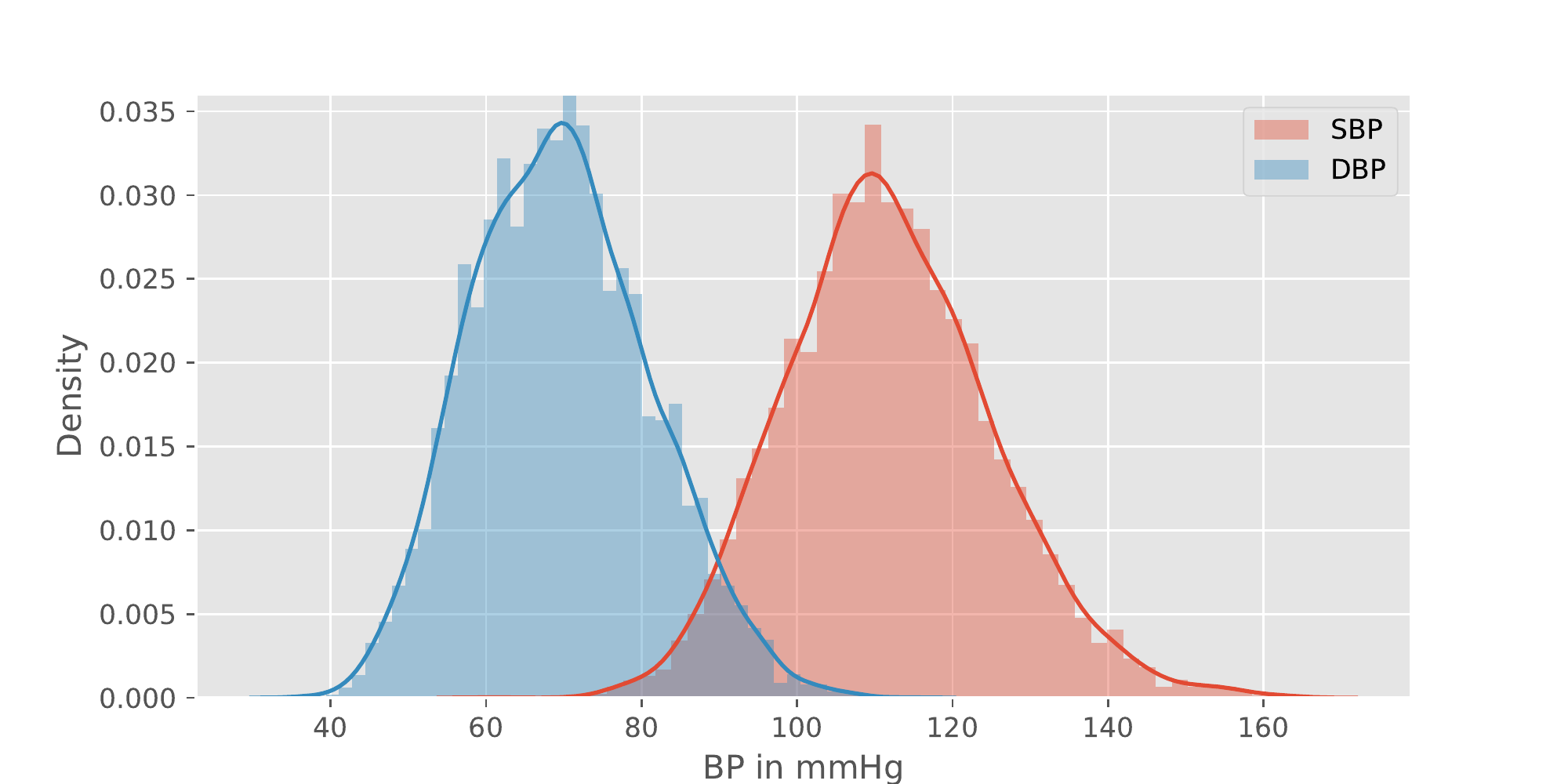}
\centering
\caption{Distribution of SBP and DBP in the data set after cleaning.}
\label{fig:dist}
\end{figure}

Data was then extracted from the devices and checked for quality. Eleven days of data were excluded due to a malfunctioning or disconnected BP monitor or less than 25\% of expected BP samples measured by the ambulatory BP monitor. In addition, measurements where SBP was $>$ 180 or $<$ 80 mmHg were regarded as noise given that the population sample was not hyper- or hypotensive. Successive differences in SBP of more than twice the SBP standard deviation were also seen as noise and discarded. This is comparable to methodology seen in \cite{Xing2016}. The remaining data consisted of 226 days from 103 participants (Age: 36.6$\pm$11.7 years, body-mass index: 24.7$\pm$4.2 $kg/m^2$, 93\% right-handed, 84\% female), including 6629 valid BP measurements extracted from the ambulatory BP monitor, which formed the ground truth for this study. The distribution of SBP and DBP in the final filtered data can be seen in Figure \ref{fig:dist}. 

\subsection{Feature extraction}
\label{sec:methods:fx}
\begin{figure*}
\centering
\includegraphics[scale=0.5]{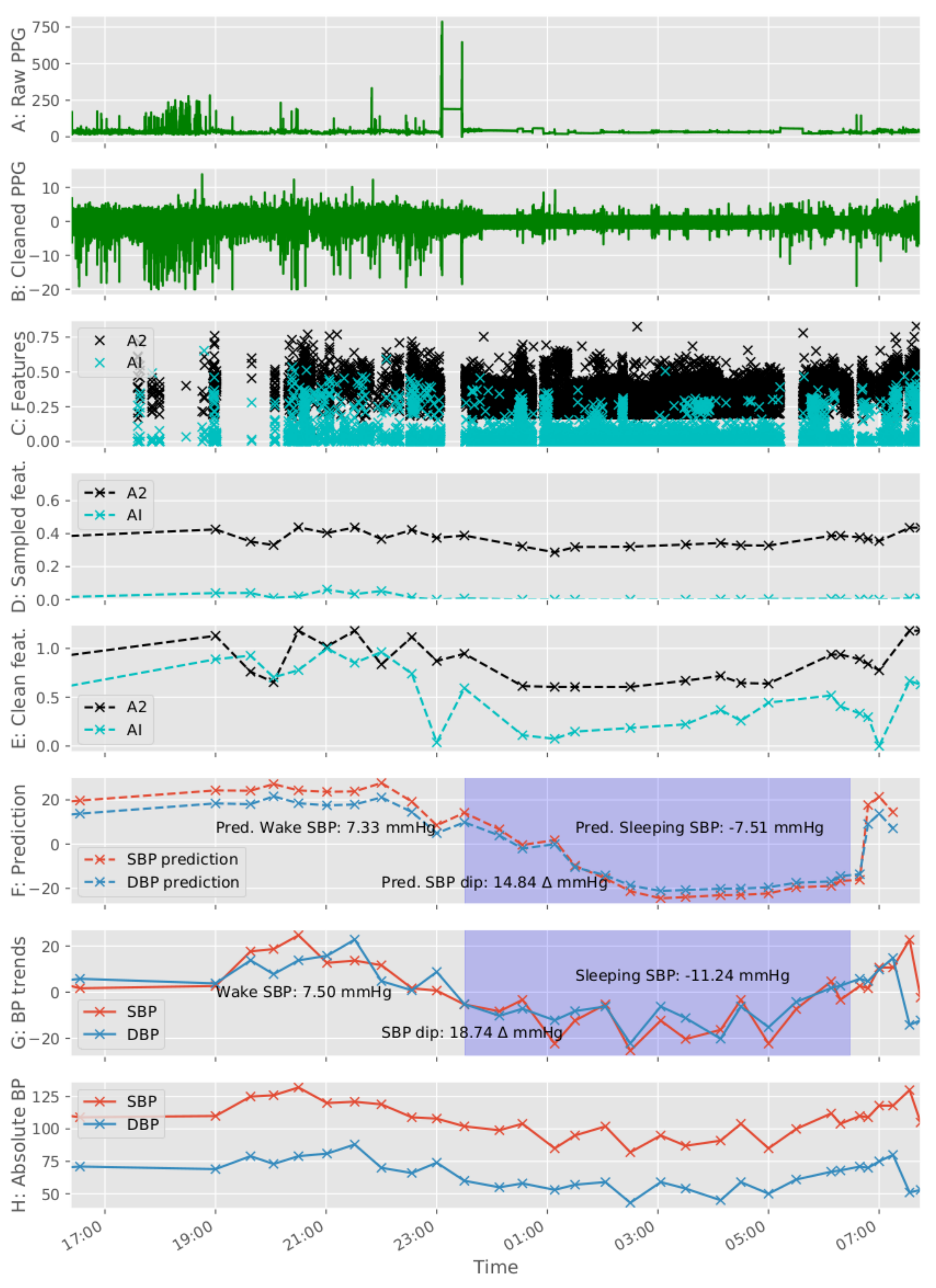}
\caption{Overview of the feature extraction procedure for one day of data (the early morning did not include many measurements and is not shown here). A: raw PPG signal; B: the filtered and down-sampled PPG signal; C: the augmentation index (AI) and Amplitude of the second pulse (A2) generated from the Gaussian mixture model features (see Figure \ref{fig:gmm} and Section \ref{sec:methods:morphology} for details); D: the same features, sampled at the same time as the ground truth was measured (see subfigure H); E: the normalised and filtered features (A2: Winsor normalisation and rolling mean filter, AI: Quantile normalization to a uniform distribution and rolling mean filter); F: the prediction of SBP by the machine learning model using all features (including AI and A2) with the sleep period highlighted, and estimation of the SBP dip from wake/sleep SBP; H (bottom Figure): the ground truth SBP and DBP; G: the normalized BP (mean removed) with sleep period highlighted and SBP dip calculated.}
\label{fig:extraction_overview}
\end{figure*}

An overview of the entire feature extraction process up to the  classification and comparison with the ground truth is given in Figure \ref{fig:extraction_overview}. 

\subsubsection{Activity features}
\label{sec:activity}
The tri-axial accelerometer data was used to estimate the likelihood of the person being at rest for non-overlapping windows of 30 seconds. It computes 1 second-based motion characteristics such as number of zero crossings, periodicity, vertical acceleration, and motion cadence. These features were aggregated into samples of 30 seconds from which the mean, standard deviation, maximum value and $95^{th}$ percentile were calculated. A pre-trained Bayesian linear discriminant (trained using a separate dataset that is described in \cite{fonseca2017validation}) was used to estimate the probability per sample that the participant was at rest in order to estimate the SBP dip (see Section \ref{sec:method:dip}). An example of a sleep-wake segmentation is shown in Figure \ref{fig:extraction_overview}F. The features were also used as an input for BP estimation.

\subsubsection{Heart rate variability}
Individual heart beats were extracted from the pulsatile component of the PPG. The same algorithm was also used by Fonseca et al. \cite{fonseca2017validation} and Chiu et al. \cite{chiu1991determination}. The beats were segmented into 5 minute windows as recommended by the heart rate variability task force \cite{camm1996heart} with an overlap of 1 minute. The variability in the beats was analyzed using the multi-scale sample entropy algorithm \cite{costa2005multiscale}, which was found to be useful for the analysis of regularity in physiological time series across different time scales. Sample entropies were calculated for scales 1 to 10, but only 6 to 10 had good correlation with SBP and thus those were the only ones used.

\subsubsection{PPG Morphology features}
\label{sec:methods:morphology}
Finally, a set of morphological features were extracted directly from the PPG pulse. The first family of features were taken from a PPG morphology review by Elgendi et al. \cite{Elgendi2012} and are based on the analysis of different peaks in the PPG pulse waveform and its derivatives.\\

\begin{figure}
\centering
\includegraphics[scale=0.55]{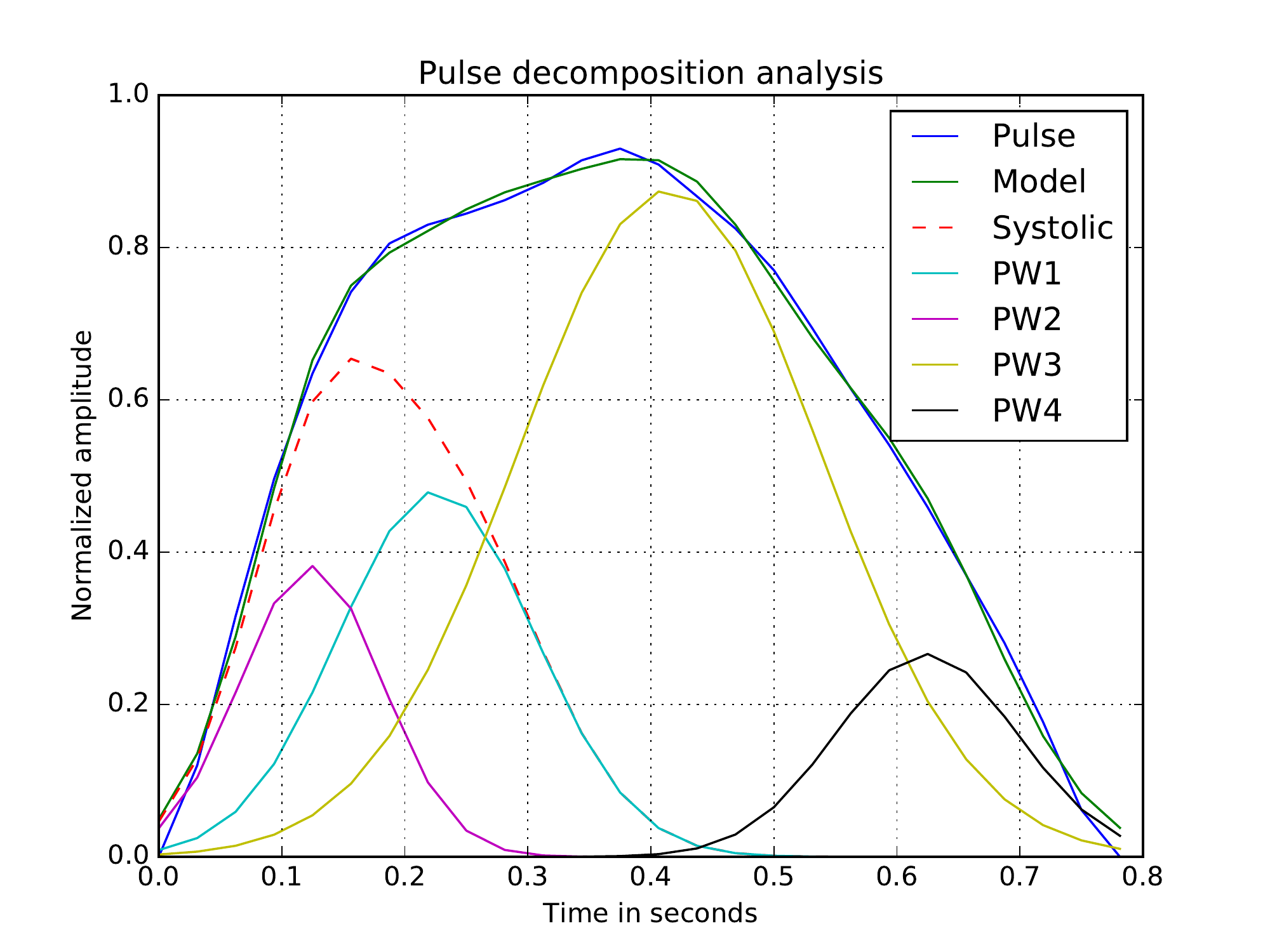}
\caption{Gaussian Mixture Model feature extraction \cite{Couceiro2012}. A single PPG pulse is decomposed into four Gaussians (PW1, PW2,PW3 and PW4), each representing different reflections in the arterial blood pulse. Amplitude, location, area and width of each pulse are used as features, as well as the difference in amplitude and area between the combined first 2 pulses (systolic pulse) and the third pulse, referred to respectively as the augmentation and reflection indices. In Figure \ref{fig:extraction_overview}C the augmentation index and the amplitude of the second pulse are used as example.}
\label{fig:gmm}
\end{figure}

The second set of features are based on pulse decomposition analysis. Each pulse wave is decomposed into a mixture of 4 gaussians \cite{Couceiro2012} that are thought to be caused by reflections in the arterial pathways due to vasoconstriction. Vasoconstriction itself is correlated with BP. The amplitudes, widths and timings of each pulsatile component were used as features. The procedure for a single PPG pulse wave is shown in Figure \ref{fig:gmm}. Two of these features were used as an example in Figure \ref{fig:extraction_overview}C to illustrate the feature extraction pipeline.\\ 

The final set of features were proposed by Monte-Moreno \cite{Monte-Moreno2011} for BP estimation. They consist of signal processing techniques to quantify entropy, irregularity and frequency content. The underlying algorithms are Shannon's entropy, Kaiser-teager energy, Qi-Zheng energy and auto-regressive analysis. The features were computed on 5 second frames of PPG with an overlap of 2.5 seconds. Subsequently, statistics were derived over all the frames per minute and used as features. These included the mean, standard deviation, inter-quartile range and skewness.\\

\subsubsection{Feature optimization}
\label{methods:feature_opt}
As discussed in the related work, while complex morphological features are known to correlate to BP in lab studies, their applicability in practice is limited. This is because they usually have a person-specific response to BP and/or may suffer from noise. Usually, features are normalized to reduce their persons-specific nature and filtered to remove noise. To determine the optimal normalizations and filters, all of the features in this study were optimized individually to maximize correlation with SBP. First, for each feature, the cross-person correlation for SBP was maximized by selecting one out of a set of normalization functions. The functions used were Z-score \cite{jain2005score}, Min-Max \cite{jain2005score}, Boxcox \cite{sakia1992box}, quantile \cite{bolstad2003comparison} and Winsor \cite{dixon1974trimming} normalization. After normalization, a set of filters was explored in the same manner to select the best filters. Considered filters were the Butterworth low-pass and high-pass filters (at a variety of cutoffs), as well as rolling mean, median and standard deviation filters. In some cases, multiple filters were found to improve a feature, and in those cases only the two best transformations were used. Optimization was done using only the training data set (see Section \ref{sec:methods:split} below). An example of the optimization of two features is shown in Figure \ref{fig:extraction_overview}.

\subsubsection{Feature extraction procedure}
\label{sec:fxp}
All features were averaged over a 5-minute window centred around each BP measurement. BP measurements where no features could be extracted in that window due to a lack of detected heart beats were excluded. When comparing a BP surrogate to a real BP signal, the low sampling frequency of the BP ground truth is not sufficient to capture high frequency changes in both BP and its predictor. Zheng et al. proposed low-pass filtering  both signals, removing frequencies above 1/30/60/3 Hertz. This technique was applied to the dataset \cite{Zheng2014}. Finally, the mean of both the targets (SBP and DBP) as well as the features per day was subtracted from the values. This allows for prediction of deviation from a basal BP level. An example of the re-sampling and filtering procedure is shown in Figures \ref{fig:extraction_overview}D,E.\\

\subsubsection{Data splitting}
\label{sec:methods:split}
After excluding invalid points (see Section \ref{sec:fxp} the final data set contained 5111 BP samples,measured using the ambulatory BP monitor ground truth and their corresponding PPG-derived feature vectors with 195 features per sample. The data of 82 randomly selected participants was reserved for training the models. The training set contained 4126 samples from 185 days. The data from the remaining 21 participants was used to evaluate the performance of the models. This test set consisted of 985 samples collected from 41 days. The training and testing set had comparable body-mass index (training: $24.3\pm3.9 kg/m^2$, testing: $24.1\pm3.5 kg/m^2$), age (training: $34.9\pm10.0$ years, testing: $36.5\pm15.4$ years), weight (training: $67.5\pm12.8$ kg, testing: $67.6\pm11.1$ kg) and sex distributions (training: 85\% female, testing: 73\% female). \\

\subsection{Models}
The following machine learning models were trained to estimate relative SBP and DBP using the created training data including: linear regression, random forest, dense neural networks and LSTM neural network.

\paragraph{Linear regression} This classic regression model trains a linear combination of all input features to generate SBP and another set of linear combinations to estimate DBP. As this model is unable to learn non-linear patterns in the data and cannot infer over temporal data, it forms the baseline machine learning model.

\paragraph{Random forest regression} Before the widespread use of neural networks, random forest models \cite{liaw2002classification} were state of the art. An ensemble of decision trees make predictions about the ground truth and the final outcome is a vote of the different trees. Random forests are very capable of inferring non-linear relationships between features and ground truth, but lack the capacity to infer over temporal structure. Three random forest regression models were trained: with 8, 16 and 32 trees.

\paragraph{Dense network} Also known as a multi-layer perceptron network \cite{ruck1990multilayer}, the dense network is the most basic of neural network architecture. It consists of perceptrons which perform linear combinations of their input (similar to linear regression) and generate intermediate outputs that are then fed to a next layer of perceptrons. The network used here consists of a first hidden layer that takes the features as input, and a second layer of perceptrons that take the first layer's outputs and predicts the BP labels. Three perceptron networks were trained using 8, 16 and 32 trees.

\paragraph{LSTM network} Finally, an LSTM model with the  potential ability to infer over the temporal structure in free-living data is trained. The LSTM cells are included in a three-layer neural network, with the first hidden layer consisting of 32 perceptrons, the second hidden layer consisting of 32 LSTM cells and the last layer consisting again of a perceptron layer to predict the labels. This model is an extension of the multilayer perceptron network network, whereby an an intermediate layer is added to allow for temporal connections to be made. The number of perceptrons in the first hidden layer was fixed at the best-performing number obtained when training the multilayer perceptron networks, while three configurations for the LSTM network were trained using 8, 16 and 32 cells.\\

Both the multilayer perceptron network and the LSTM network are trained through back-propagation with a batch size of 32 days of data at a time. Of the training data set, 80\% was used in back-propagation while the remainder was used to see whether the model was converging. To avoid overfitting, once the loss on this subset stopped improving for 20 consecutive passes over the data, the training was stopped (early stopping criterion \cite{prechelt1998early}). \\ 

For all machine learning models, relative mean arterial pressure was also included as a third output of the model. This was done to improve convergence of the more complex models such as the random forest and two neural networks (same approach as introduced in \cite{su2017predicting}). The loss function used for all models is the mean squared error. However, as mean squared error optimization favours minimizing instances with large errors, the criterion is known to lead to models that are not sensitive to outliers. This is a challenge for regression models, especially ones where the ground truth is normally distributed such as in this data set (see Figure \ref{fig:dist}). To remedy this, the loss was amplified by the absolute difference of the SBP value from the mean SBP in the training set. This means that the larger the deviation of the SBP value from the mean, the more its loss was amplified. This amplification remedied the outlier insensitivity in \textit{all} of the machine learning models.

\subsection{Blood pressure tracking evaluation}
\label{sec:method:dip}
The quality of the predictions on the test set were compared using two criteria: the RMSE as well as the Pearsons' correlation between ground truth and predicted SBP/DBP. Both metrics were calculated per day and both the mean and standard deviation are reported.\\

\subsection{Blood pressure dip evaluation}
The use case for the developed models is to estimate the SBP dip as a daily SBP statistic from the trend estimations. The procedure is illustrated in Figures \ref{fig:extraction_overview}F,G. The day is segmented into either sleep or wake segments. The dip is then calculated as the difference between the mean SBP during the wake segment and the mean SBP during the sleep segment. Note that this procedure \textit{does not require recalibration} to absolute values, as the dip itself is a relative measure. The predicted and true SBP dip are also compared through RMSE and correlation analysis. The prediction performance of the machine learning models will be compared to two base line models that do not rely on machine learning: 

\paragraph{Demographic-based SBP dip estimation} A linear regression model of the SBP dip will be fitted on the training data using only age, weight and body-mass index characteristics of participants and evaluated on the test set as a baseline that does not require physiological signals.

\paragraph{Heart rate based SBP dip estimation} Heart rate also exhibits a dipping pattern controlled partially by the same physiological mechanisms as the SBP dip \cite{fava2005dipping,sherwood2002nighttime}. The dip in the heart rate will be computed in the same manner as for the SBP dip (difference between mean heart rate during the day and mean heart rate at night) and a linear fit will be determined between the heart rate and SBP dips using the training set. This will form a second baseline that does use physiological signals, but does not require a model to estimate SBP first.

\section{Results}
\label{sec:results}
\begin{table}
\resizebox{\textwidth}{!}{
\begin{tabular}{l|ll|ll|ll}
& \multicolumn{2}{c|}{\textbf{Relative SBP}}& \multicolumn{2}{c|}{\textbf{Relative DBP}}& \multicolumn{2}{c}{\textbf{SBP dip}}         \\
& \multicolumn{1}{c}{\textbf{\begin{tabular}[c]{@{}c@{}}RMSE\\ (mmHg)\end{tabular}}} & \multicolumn{1}{c|}{\textbf{\begin{tabular}[c]{@{}c@{}}Corr\\ (Pearson)\end{tabular}}} & \multicolumn{1}{c}{\textbf{\begin{tabular}[c]{@{}c@{}}RMSE\\ (mmHg)\end{tabular}}} & \multicolumn{1}{c|}{\textbf{\begin{tabular}[c]{@{}c@{}}Corr\\ (Pearson)\end{tabular}}} & \multicolumn{1}{c}{\textbf{\begin{tabular}[c]{@{}c@{}}RMSE\\ (mmHg)\end{tabular}}} & \multicolumn{1}{c}{\textbf{\begin{tabular}[c]{@{}c@{}}Corr\\ (Pearson)\end{tabular}}} \\
\rowcolor[HTML]{EFEFEF} 
\rowcolor[HTML]{EFEFEF}\textbf{Linear regression}& 8.90$\pm$2.11& 0.68$\pm$0.16& 6.76$\pm$1.48& 0.76$\pm$0.13& 3.92$\pm$3.36& $0.49^{*}$\\

\textbf{\begin{tabular}[c]{@{}l@{}}Random forest\\  (8 trees)\end{tabular}}   & 8.20$\pm$1.50& 0.64$\pm$0.16& 7.03$\pm$1.25& 0.72$\pm$0.13& 4.46$\pm$2.30& $0.61^{***}$\\

\rowcolor[HTML]{EFEFEF} 
\textbf{\begin{tabular}[c]{@{}l@{}}Random forest \\ (16 trees)\end{tabular}}  & 7.90$\pm$1.46& 0.69$\pm$0.14& 6.68$\pm$1.44& 0.75$\pm$0.13& 4.65$\pm$3.00& $0.63^{***}$\\

\textbf{\begin{tabular}[c]{@{}l@{}}Random forest\\  (32 trees)\end{tabular}}  & \textbf{7.86}$\pm$\textbf{1.57}& 0.69$\pm$0.16& 6.75$\pm$1.41& 0.75$\pm$0.13& 4.75$\pm$3.33& $0.57^{**}$\\

\rowcolor[HTML]{EFEFEF}
\textbf{\begin{tabular}[c]{@{}l@{}}Dense \\ (8 perceptrons)\end{tabular}}  & 9.07$\pm$2.01& 0.71$\pm$0.15& \textbf{6.49}$\pm$\textbf{1.59}& 0.77$\pm$0.12& 4.68$\pm$3.43& $0.66^{****}$\\

\textbf{\begin{tabular}[c]{@{}l@{}}Dense \\ (16 perceptrons)\end{tabular}} & 9.21$\pm$2.09& 0.71$\pm$0.15& 6.53$\pm$1.56& 0.77$\pm$0.12& 4.70$\pm$3.46& $0.65^{****}$\\

\rowcolor[HTML]{EFEFEF} 
\textbf{\begin{tabular}[c]{@{}l@{}}Dense \\ (32 perceptrons)\end{tabular}} & 9.27$\pm$2.09& 0.71$\pm$0.15& 6.53$\pm$1.52& 0.77 $\pm 0.12$& 4.52$\pm$3.41& $0.65^{****}$\\

\textbf{\begin{tabular}[c]{@{}l@{}}LSTM \\ (8 cells)\end{tabular}}            & 8.12$\pm$1.67& 0.70$\pm$0.12& 6.68$\pm$1.37& 0.76$\pm$0.12& 3.56$\pm$2.28& $0.63^{***}$\\

\rowcolor[HTML]{EFEFEF} 
\textbf{\begin{tabular}[c]{@{}l@{}}LSTM\\ (16 cells)\end{tabular}}            & 7.96$\pm$1.63& 0.69$\pm$0.17 & 6.57$\pm$1.51& 0.76$\pm$0.13& 3.45$\pm$2.33& $0.63^{***}$\\

\textbf{\begin{tabular}[c]{@{}l@{}}LSTM\\ (32 cells)\end{tabular}}            & 8.22$\pm$1.49& 0.69$\pm$0.16& 6.55$\pm$1.39& 0.76$\pm$0.13  & \textbf{3.12}$\pm$\textbf{2.20}& \textbf{0.69}$^{****}$           \\         
\end{tabular}}
\caption{Overview of the performance of the models evaluated in this study. The number of stars in the last column denotes the significancy level of the correlation: *, **, *** and **** respectively denote $p < 0.05, p<10^{-2}, p<10^{-3}$ and $p<10^{-4}$. RMSE, correlation (Corr) for SBP and DBP were computed per day of data and reported as the mean over all days in the test set $\pm$ the standard deviation. Numbers in bold are the highest scores for the metric. For correlation of SBP and DBP the numbers were too close to select a clear best score and thus none of the entries are in bold.}
\label{tab:performances}
\end{table}

The main results of the study are included in Table \ref{tab:performances}. The performance is shown for ten models: linear regression, random forest models with 8, 16 and 32 trees, dense networks with 8, 16 and 32 perceptrons and finally LSTM networks with 8, 16 and 32 cells. For the LSTM models, the first hidden layer had 8 dense perceptrons as the performance of all three dense networks was close to each other. RMSE and correlation were computed per day of data and both averages and standard deviations are given in Table \ref{tab:performances}. For the SBP dip estimation task, the sleep-wake classifier was used to segment every day into waking and sleep. The difference between mean SBP during the wake and sleep periods was calculated per day. The RMSE for the estimated SBP dips is also given in Table \ref{tab:performances}. Correlation was computed once for the entire test set (using every day as a sample) and the significance level is reported in the table.\\

Next to the machine learning models, two baseline models were also tested. The linear regression model based on demographics was trained using age, sex and body-mass index as input variables on the training set and evaluated on the test set. It had an RMSE of 2.22$\pm$5.36 $\Delta$ mmHg and a non-significant correlation of 0.15 ($p=0.44$) with the ground truth. The baseline model based on heart rate was not trained. Instead, the dip in heart rate was computed in the same manner as was done for SBP and a correlation analysis between the heart rate and SBP dip was performed, resulting in a negative non-significant correlation of -0.27 ($p=0.16$) with the ground truth.\\

\begin{figure}
\centering
\begin{subfigure}{.49\textwidth}
  \centering
  \includegraphics[scale=.49]{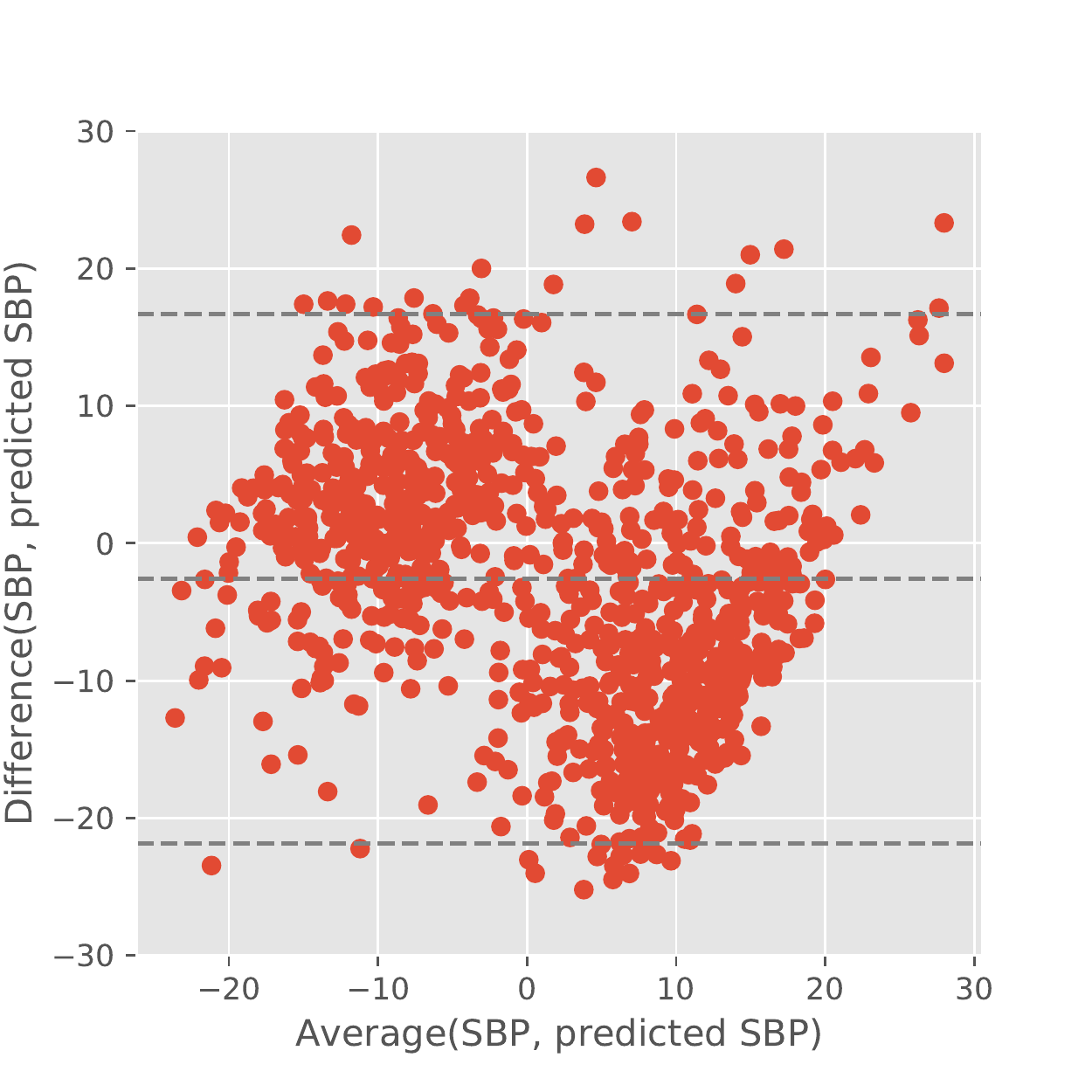}
  \caption{Relative SBP, using LSTM with 32 cells\centering}
  \label{fig:ba_sbp_lstm}
\end{subfigure}%
\vspace{\floatsep}
\begin{subfigure}{.49\textwidth}
  \centering
  \includegraphics[scale=.49]{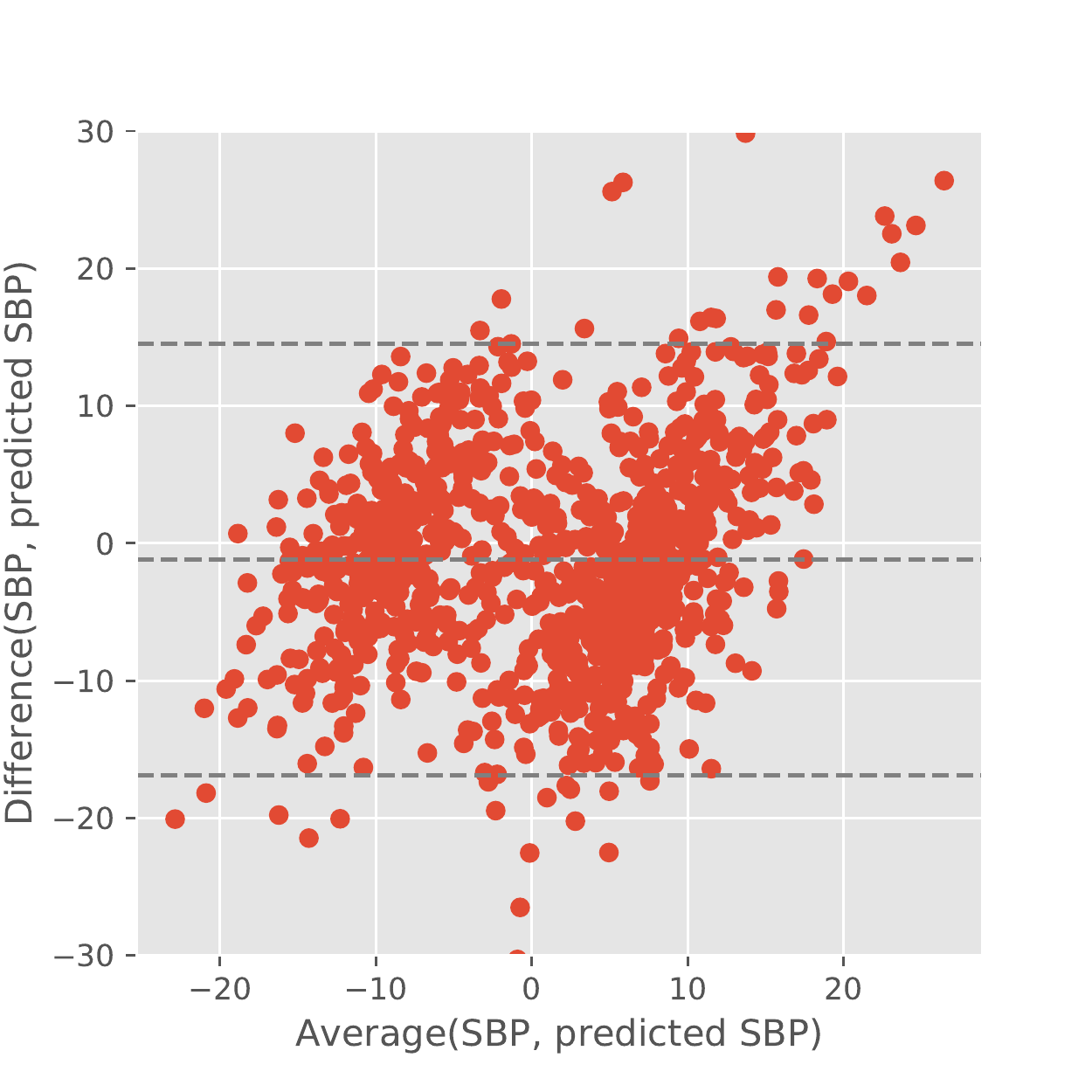}
  \caption{Relative SBP, using random forest with 32 trees\centering}
  \label{fig:ba_sbp_rf}
\end{subfigure}%
\vspace{\floatsep}
\begin{subfigure}{.49\textwidth}
  \centering
  \includegraphics[scale=.4]{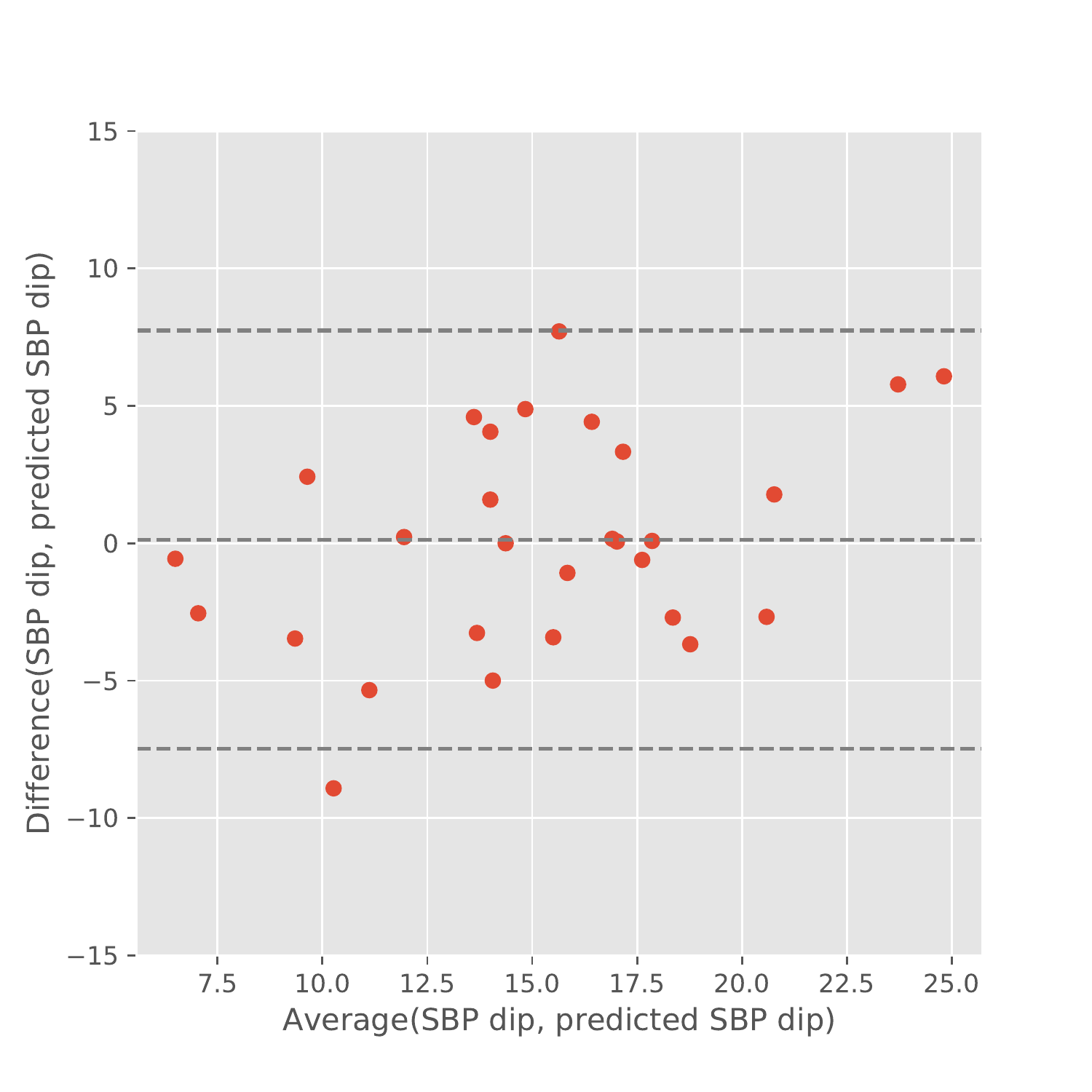}
  \caption{SBP dip, using LSTM with 32 cells\centering}
  \label{fig:ba_dip_lstm}
\end{subfigure}%
\begin{subfigure}{.49\textwidth}
  \centering
  \includegraphics[scale=.4]{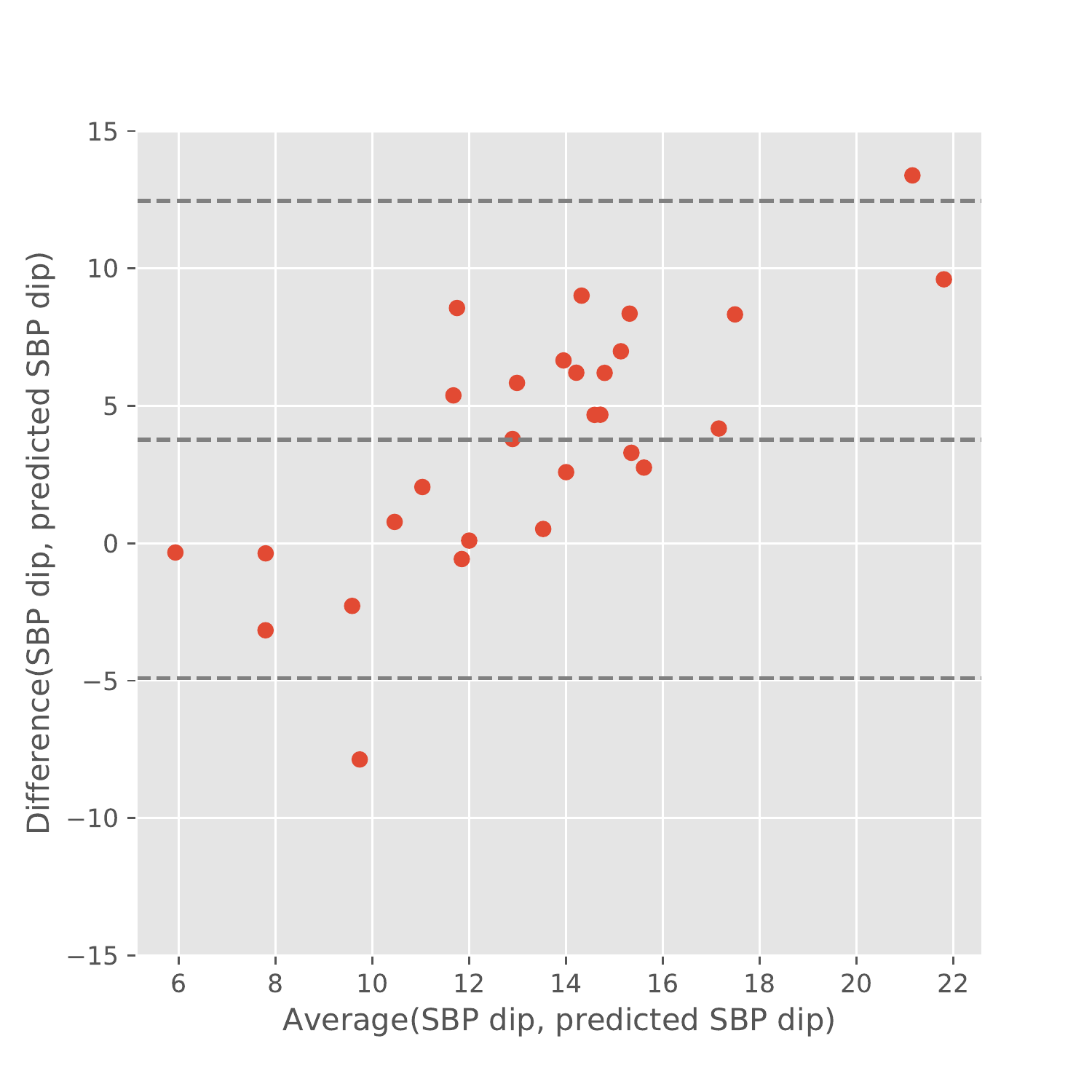}
  \caption{SBP dip, using random forest with 32 trees\centering}
  \label{fig:ba_dip_rf}
\end{subfigure}
\caption{Comparison of systematic errors between the LSTM and random forest models through Bland-Altman analysis of the uncalibrated values of SBP (upper subfigures) as well as the SBP dip (lower subfigures).}
\label{fig:ba_analysis}
\end{figure}

A Bland-Altmann analysis was performed to compare the best-performing model for SBP prediction (random forest with 32 trees) to the best-performing model for SBP dip prediction (LSTM with 32 cells) on their ability to estimate both the SBP and the SBP dip, as visualized in Figure \ref{fig:ba_analysis}. The LSTM model had the following estimation errors (mean$\pm$standard deviation): -2.60$\pm$9.82 mmHg for SBP and 0.13$\pm$3.88 mmHg for the SBP dip. The random forest model had an estimation error of -1.17$\pm$8.01 for SBP and 3.77$\pm$4.43 for the SBP dip.

\begin{figure}
\centering
\begin{subfigure}{.49\textwidth}
  \centering
  \includegraphics[scale=.45]{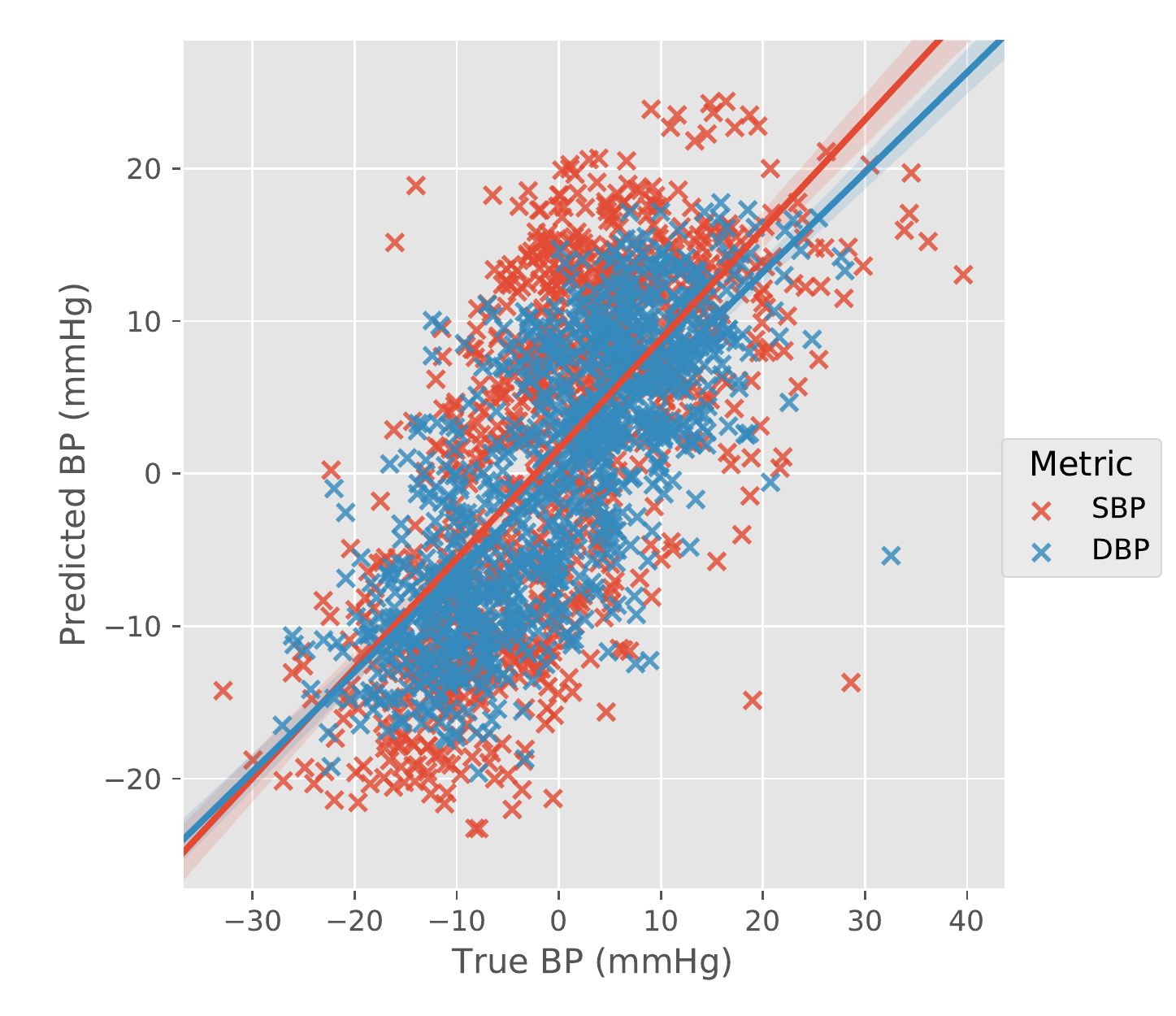}
  \caption{Relative SBP and DBP\centering}
  \label{fig:sbp_pred}
\end{subfigure}%
\vspace{\floatsep}
\begin{subfigure}{.49\textwidth}
  \centering
  \includegraphics[scale=.4]{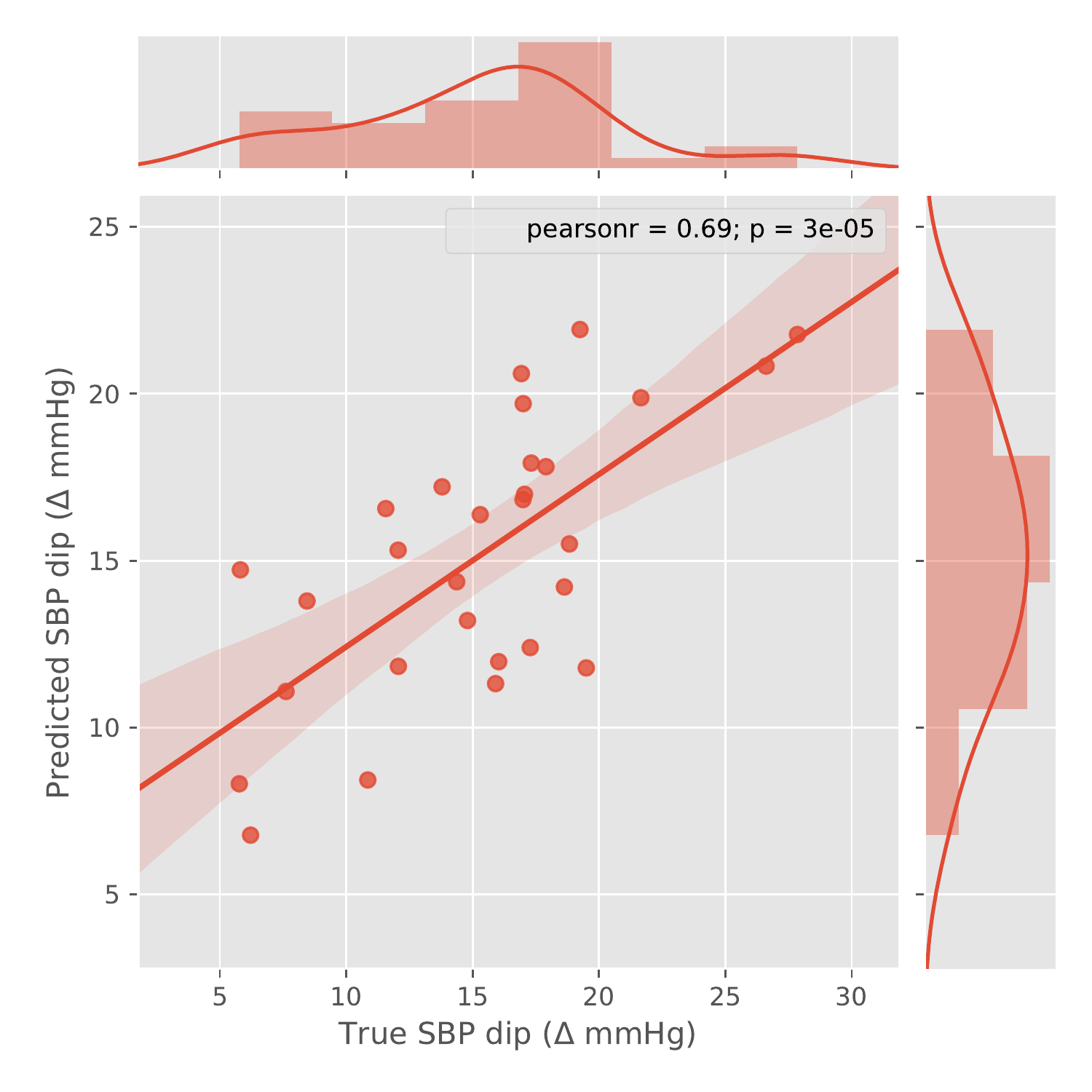}
  \caption{SBP dip\centering}
  \label{fig:dip_pred}
\end{subfigure}
\caption{Correlation and regression analyses for the prediction of relative SBP and DBP (left) and the SBP dip (right) using the LSTM with 32 cells.}
\label{fig:correlation_analyses}
\end{figure}

Finally, regression and correlation analyses for the estimation of SBP and DBP using the LSTM model with 32 cells are visualized in Figure \ref{fig:sbp_pred} while the same analyses for the model's estimations of the SBP dip are shown in Figure \ref{fig:dip_pred}.

\section{Discussion}
\label{sec:discussion}
A system for the estimation of BP trends was evaluated in 24-hour free-living individuals, with two main evaluations: the accuracy of estimating relative changes in  SBP and DBP, as well as the accuracy of the nocturnal SBP dip estimates. The system used PPG and acceleration from a wearable wrist device as input and features were extracted based on heart rate variability and pulse morphology. Several machine learning models were compared in their accuracy. In this section the predictive capacity of the models is discussed and compared to earlier studies. The limitations are highlighted and suggestions for future work to improve accuracy and clinical applicability is proposed.

\subsection{Machine learning models}
\label{sec:dicussion:ml}
Several models representing different categories were evaluated. This included linear regression which represents a strictly linear model, the random forests and dense networks which represent non-linear models while the LSTM model represents a sequential model (both non-linear and capable of inference over temporal structure in the data). When comparing the 
performance of the estimates of these models with the ground truth for SBP and DBP tracking, they all have a similar level of accuracy. For SBP however, the random forest with 32 trees and the LSTM with 8 cells had respectively the best RMSE and correlation, though the differences with other models were minimal and not significant. For DBP prediction, the dense model with 8 perceptrons achieved the best performance, but again it was very close to the other models.\\

The results become more informative when considering the SBP dipping estimations. The LSTM with 32 cells achieved the highest correlation between its estimate of the dip and the true dip of 0.69 ($p=3*10^{-5}$), while the linear regression model achieved the lowest with a correlation coefficient of only 0.49 ($p=0.03$), providing the first evidence for the superiority of sequence-to-sequence modelling over linear modelling. All other machine learning models also strongly outperformed the baselines. Both heart rate based and demographics-based dip estimation did not correlate with the true SBP dip. For all other models, the correlation ranged between 0.57 and 0.65, which is still markedly lower numbers than for the LSTM model with 32 cells, especially in comparison to the limited differences between the models in estimation accuracy of relative SBP and DBP.\\ 

This is likely because the estimation of the dip not only requires the direction of the BP trend to be estimated, but also the \textit{magnitude of change}. This becomes more clear in the Bland-Altmann analysis presented in Figure \ref{fig:ba_analysis}. The errors for the estimation of the relative SBP values are similar for both the random forest (Figure \ref{fig:ba_sbp_rf}) and LSTM models (Figure \ref{fig:ba_sbp_lstm}), however for the SBP dip (Figures \ref{fig:ba_dip_lstm} and \ref{fig:ba_dip_rf}) there is a clear difference. As the dip assessment relies on good approximation of deviation from the mean, the random forest model shows both an over-estimation bias as well as a strong amplitude-dependent error while the errors of the LSTM model is not biased and shows a smaller amplitude-dependent error. This implies that even though the RMSE for SBP was lower for the random forest model, the LSTM model still has a more consistent prediction capability because of a better sensitivity to BP values that are deviant from the mean. 

\subsection{Comparison to earlier work}
In Section \ref{sec:background}, related work was discussed and summarized in Table \ref{tab:state_of_art}. Most of the studies presented in this table (apart from Zheng et al. \cite{Zheng2014} and Ruiz-Rodriguez et al. \cite{Ruiz-Rodriguez2013}) have been performed in controlled lab studies, in which effects as described in Section \ref{sec:dicussion:ml} could not be tested for. The studies involving point measurements had the highest accuracy level, but the application areas of such methods is limited as point measurements can also be performed with a regular BP cuff. Studies involving continuous measurements might at first sight give the impression of being more rigorously tested than point measurements, but machine learning models can overfit to the specific experimental protocol with which the data is obtained. For example, in a highly controlled intermittent exercise test on an ergometer, the increase in heart rate will strongly relate to the increase in BP as they are both primarily being controlled by oxygen demands in the muscles. This relation may be exploited by a machine learning model to generate seemingly accurate results. However once a person is monitored outside of exercise, the correlation between heart rate and BP weakens, making such a model invalid. The weak correlation between heart rate and BP was observed in this study as well: the heart rate based dip estimation had no correlation with the BP dip. The lower performance achieved in this work for PPG-based monitoring in comparison to lab studies that controlled lab-based evaluation of unobtrusive BP estimation might suffer from poor external validity as results cannot be extrapolated to real-world use cases.\\

Zheng et al. \cite{Zheng2014} was the only work to evaluate their method in free-living individuals. However the RMSE was only reported for the night-time period and the report discloses that day-time PAT has a very limited correlation with BP due to its high sensitivity to posture changes. Thus the method is not suitable for the estimation of the SBP dipping for which day-time BP is also needed. In that work, PAT was esimated through a unique technology in which both ECG and PPG could be measured with a single wrist device. A future study in which the ECG-to-PPG measure of PAT from such a device could be combined with heart rate variability and PPG morphology as used in this work can lead to marked improvements in the accuracy of the method with minimal impact on the comfort of the wearer, especially during the night time.\\

 Ruiz-Rodriguez et al. \cite{Ruiz-Rodriguez2013} used a large cohort of intensive care patients. The reported error margins are slightly higher than in the current study, which could be because they estimated absolute BP or because intensive care patients may suffer from extreme health conditions that lead to inconsistent pulse morphology. However, the low mobility of patients in intensive care units does eliminate confounding factors related to motion and posture change. In short, the protocols used by Ruiz-Rodriguez et al. and this study are highly disparate making it difficult to compare performance. Future work in which a model is trained on both clinical intensive care data as well as free-living individuals would likely result in an algorithm that can handle a very wide range of use cases.\\

\subsection{Sources of error and limitations}
The results presented in this study provide encouraging evidence for BP trend measurement using PPG in free-living individuals. However there are a few limitations that could be addressed in future work to improve the model's performance. One likely source of error could come from motion artefacts as free-living individuals are not restricted in their activities. Such artefacts also reduce the coverage of the method, as low-quality PPG data is automatically discarded by the feature extraction methodology we used. One possible solution to this is to use a PPG sensor mounted to a body part that undergoes less movement than the wrist such as the chest or head. However, although a different body location may suffer less from motion artefacts, there would be the additional challenge of fixing a PPG sensor to those body parts. At least with the wrist an occlusive band can be tightened to the right skin contact level, without requiring any adhesives or causing discomfort.\\

Another limitation for this study is the ambulatory BP monitor ground truth. These monitors are highly obtrusive and can only measure semi-continuously. During sleep their obtrusiveness becomes close to unbearable (which was a motivation for the method proposed in this study in the first place). Other studies have used non-invasive continuous BP monitors based on the volume clamp principle \cite{Ruiz-Rodriguez2013,su2017predicting} which yields a high temporal resolution. However this method is already on its own a surrogate for BP estimation and is not recommended for clinical use \cite{berne1967cardiovascular}. Moreover, Imholz et al. \cite{imholz1993feasibility} showed that using the volume clamp principle suffers from structural under-estimations in daily life activities. A far better ground truth is the invasive BP estimation procedure through catheters, but patients undergoing such treatments are rare, not mobile and often suffer from complicated pathologies, making them a difficult-to-access population for large-scale studies. Nevertheless data from both these sensors could be used as a supplementary source to pre-train the model, which can then be adapted to free-living data through techniques such as transfer learning \cite{pan2010survey}. \\

\subsection{Clinical significance}
Finally, a note on the clinical significance of the results should be made. While the used data set in this study is larger than most previous work and the results provide a first view about the utility of PPG morphology in free-living individuals,  only healthy individuals were included. The next research step would be to evaluate the methodology used in this study on populations for which such technology may provide actionable information for clinical treatment. An example of this would be sleep apnea patients who are known to suffer from non-dipping \cite{loredo2001sleep}. The use of the proposed SBP dip estimation method in early screening and home assessment procedures could potentially help clinicians to more efficiently triage patients. Another use case is in occupational health of regular shift-workers where a disruption in the circadian rhythm can lead to (temporary) shifts in the SBP diurnal rhythm that may manifest as an unhealthy dipping pattern \cite{sternberg1995altered}. Finally, regular screening for hypertension usually only involves measuring day-time point measurements as the 24-hour ambulatory measurement procedure is too burdensome to prescribe to all patients as a screening tool. Disrupted night-time BP may be overlooked and underdiagnosed because of this.  The proposed method is a highly ergonomic surrogate to the cuff-based ambulatory procedure that could be used to screen for non-dipping and thus allow for the appropriate prescription of chrono-therapy to help control this risk factor of cardiovascular disease, as evidenced in the work of Hermida et al. \cite{hermida2008chronotherapy}.

\section{Conclusions}
\label{sec:conclusions}
Several machine learning models have been compared for the prediction of relative SBP and DBP in free-living as well as the SBP dip. The methods were evaluated in a large cohort of 103 participants measured for a combined total of 226 days. The neural network consisting of 8 perceptrons in the first hidden layer and 32 LSTM cells in the second hidden layer was found to produce best results for the estimation of the SBP dip. It had the highest Pearsons' correlation (0.69, $p=3*10^{-5}$) and lowest error (RMSE of 3.12$\pm$2.20 mmHg) for the estimation of the SBP dip. For tracking of relative SBP and DBP there were no big differences in RMSE or correlation between random forests, dense networks or LSTM networks. However, further analysis revealed that the LSTM model was better at predicting deviant values of SBP. Thus the LSTM model was overall the best model.\\

The reported error rates are higher than what is typically measured in lab studies, even though the used algorithms were comparable to those of other PPG-only studies. This warrants that performance results of BP surrogates obtained in strongly controlled environments may suffer from poor external validity for real-world usage.

\section{Acknowledgements}
The authors would like to thank the European Institute of Technology for funding the project and the many involved individuals from Imperial College London and Royal Philips. Special thanks to Prof. Azeem Majeed, Dr. Josip Car, Dr. Antonio Vallejo-Vaz and Prof. Kausik K. Ray for their support and insights in the organisation and clinical judgements of the study.

\section*{Bibliography}

\bibliographystyle{iopart-num}
\bibliography{library.bib}

\end{document}